**Data mining of public genomic repositories: harnessing off-target reads to expand microbial pathogen genomic resources**

Damien Richard*[$,1], Nils Poulicard[$,2]

* Corresponding author, damien.richard@ird.fr
[$] PHIM Plant Health Institute, Université Montpellier, IRD, INRAE, Cirad, Institut Agro, Montpellier, France
[1] ORCID: 0000-0001-9179-7315
[2] ORCID: 0000-0003-3969-2131

**Abstract**

As sequencing technologies become more affordable and genomic databases expand continuously, the reuse of publicly available sequencing data emerges as a powerful strategy for studying microbial pathogens. Indeed, raw sequencing reads generated for the study of a given organism often contain reads originating from the associated microbiota. This review explores how such off-target reads can be detected and used for the study of microbial pathogens. We present genomic data mining as a method to identify relevant sequencing runs from petabase-scale databases, highlighting recent methodological advances that allow efficient database querying. We then briefly outline methods designed to retrieve relevant data and associated metadata, and provide an overview of common downstream analysis pipelines. We discuss how such approaches have (i) expanded the known genetic diversity of microbial pathogens, (ii) enriched our understanding of their spatiotemporal distribution, and (iii) highlighted previously unrecognized ecological interactions involving microbial pathogens. However, these analyses often rely on the completeness and accuracy of accompanying metadata, which remain highly variable. We detail common pitfalls, including data contamination and metadata misannotations, and suggest strategies for result interpretation. Ultimately, while data mining cannot replace dedicated studies, it constitutes an essential and complementary tool for microbial pathogen research. Broader utility will depend on improved data standardization and systematic genomic monitoring across ecosystems.

**Introduction**

As sequencing goes down in price, public genomic databases go up in size (Sayers et al., 2023). The most widely used databases are the DNA Data Bank of Japan, GenBank, and the European Nucleotide Archive, all involved in the International Nucleotide Sequence Database Collaboration. They include assembled genomes or metagenomes, but also sequencing reads from which the assemblies are derived. These sequencing reads have sometimes not yet fully been explored outside of the scientific context for which they were generated. Notably, pathogens are often represented in off-target reads of studies initially designed to sequence their host or their host's transcriptome in which a given pathogen may have gone unnoticed due to symptomless infections, co-infection, or lack of interest/knowledge from the researchers and institutions that initiated the sequencing effort. Specificities in the biological origin (lab-grown vs. wild organism), wet-lab methods (tissue-specific sequencing, enrichment or purification steps…) or type of sequencing (Whole Genome, Exome, RNA, Amplicon, or Single-Cell Sequencing for example) influence the presence and relevance of such off-target reads for a given pathogen. The wealth of unexplored data present in raw read databases is increasingly leveraged by researchers, whether to complement their own datasets or to conduct entirely database-based studies (Kawasaki et al., 2023; Lagzian et al., 2024). This source of data is interesting to explore as it supplements available genomes, which are frequently biased because they were generated in response to disease outbreaks and not as part of continuous monitoring programs. For molecular pathologists, a straightforward approach consists of re-analysing all available sequencing runs corresponding to the host(s) infected by the pathogen they study (Jones et al., 2025). However, more efficient, exhaustive, and assumption-free approaches also exist. In this review, we focus on case studies of microbial pathogens that involve the screening of genomic datasets from NCBI databases (Sayers et al., 2025). We briefly review the existing methods associated with genomic data mining and show some of the benefits and limitations of using publicly available genomic resources for the study of microbial pathogens.

**Methodological aspects of genomic data mining**

Genomic data mining, i.e. here referring to the process of identifying sequencing projects of interest among databases, poses a series of challenges, the greatest of which is technical and lies in the amount of data to be screened to find relevant sequencing runs (in February 2024, the Short Read Archive (SRA) database contained 53 petabases, i.e. $5.3 \times 10^{16}$ bases). Recent methodological and computational advances, as well as the demand for genomic resources to characterize the recent SARS-CoV-2 pandemic, have fuelled the development of methods aimed at indexing large-scale sequence databases (Edgar et al., 2022; Karasikov et al., 2024; Katz et al., 2021; Shiryev & Agarwala, 2024). While building the index is computationally demanding, querying it is very efficient. Depending on the indexing method and the study case, the query can be a sequence, (be it a transcript or an antimicrobial gene, for example), or a taxon (a pathogen, a host, or a vector). Research groups developing these indexing methods sometimes precompute database indexes and make them available online to the research community (Table 1). This pre-filtering step offers to the researcher the advantage of not having to download and process all the sequencing runs, but rather of focusing computing resources on the analysis of a relevant subset. The downside of using such online query tools is that the index does not always contain the entirety of the SRA database nor an up-to-date version of it, and that queries are sometimes limited to taxa present in GenBank RefSeq (Table 1). The choice of the data mining method must be informed and mainly depends on (i) the target database (metagenomes, human, all SRA), (ii) the query organism, (iii) the availability of a RefSeq genome (referring to a GenBank reference sequence) and (iv) the research question addressed. Regardless of the tool used, one should end up with sequencing run accessions matching the query, which can be further downloaded and processed locally. In summary, the research community can now query petabase-scale databases in search for sequencing runs useful to address its research interests.

Table 1. Overview of tools designed to screen sequencing runs in public genomic databases

| Tool | Publication | Nature of the query | Pre-computed database(s) | Online access |
| --- | --- | --- | --- | --- |
| STAT | Katz et al., 2021 | Taxon (NCBI RefSeq taxid1) | SRA (all) | https://www.ncbi.nlm.nih.gov/sra/docs/sra-taxonomy-analysis-tool/ |
| Metagraph | Karasikov et al., 2024 | Sequence | SRA-Fungi, SRA-MetaGut, SRA-Metazoa, SRA-Microbe, SRA-Mouse | https://metagraph.ethz.ch/search/sra_metagut |
| Pebblescout | Shiryev & Agarwala, 2024 | Sequence | SRA microbe, SRA metagenomic | https://pebblescout.ncbi.nlm.nih.gov/ |
| Serratus | Edgar et al., 2022 | RNA virus family, RefSeq GenBank accession2, or SRA run identifier3 | SRA (all) | https://serratus.io/explorer/ |

1. NCBI RefSeq taxid refers to the unique taxonomy identifier assigned by the NCBI Reference Sequence (RefSeq) database for standardized classification of organisms.
2. RefSeq GenBank accession refers to the unique identifier assigned to a reference genomic, transcript, or protein sequence in the NCBI GenBank database, as part of the curated RefSeq collection.
3. SRA run identifier refers to a unique ID assigned to a specific sequencing run in the NCBI Sequence Read Archive (SRA), representing raw sequencing data from a single experiment.

**Downstream analyses of obtained sequencing data**

Processing the sequencing runs identified by the data mining approach usually relies on a series of bioinformatic steps. These include database-specific methods to download the sequences and the associated metadata, but also approaches commonly used in Next-Generation Sequencing (NGS) and metagenomics studies such as mapping-based variant calling, *de novo* assembly, or taxonomic assignation. Sequencing runs can be downloaded from NCBI SRA using the SRA Toolkit (https://github.com/ncbi/sra-tools). The associated metadata are stored in the dedicated BioSample database which can be interrogated manually (https://www.ncbi.nlm.nih.gov/biosample) or programmatically (NCBI E-utilities https://www.ncbi.nlm.nih.gov/books/NBK179288/, ffq tool; Gálvez-Merchán et al., 2023) for larger datasets. Next, in the context of microbial pathogens, subsequent analyses can include (i) spatiotemporal and ecological distribution by metadata analysis, (ii) phylogenetics by reference-based study, (iii) structural genomics through *de novo* assembly, and/or (iv) pathobiome community characterization. Each of these requires multiple analytical steps (Figure 1).

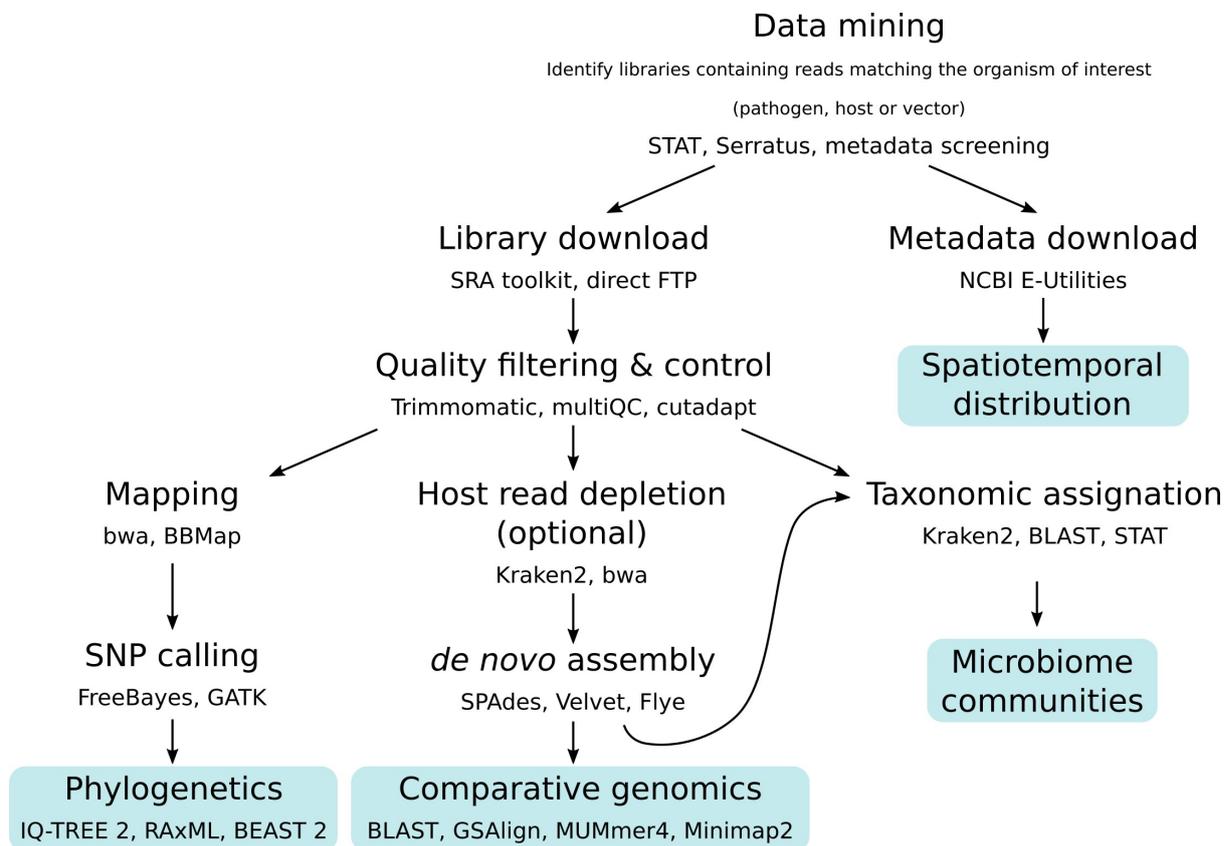

Figure 1. Schematic of simplified possible analysis pipelines following a genomic data mining approach. Tools provided as examples include STAT (Katz et al., 2021) and Serratus (Edgar et al., 2022) for data mining; SRA toolkit (National Center for Biotechnology Information, 2024) for library download; NCBI E-Utilities (Kans, 2013) for metadata download; Trimmomatic (Bolger et al., 2014), multiQC (Ewels et al., 2016) and cutadapt (Martin, 2011) for quality filtering and control; Kraken2

(Wood et al., 2019) and bwa (Li & Durbin, 2009) for host read depletion; bwa (Li & Durbin, 2009), BBMap (Bushnell, 2014) for mapping; Kraken2 (Wood et al., 2019), BLAST (Altschul et al., 1990), and STAT (Katz et al., 2021) for taxonomic assignation; FreeBayes (Garrison & Marth, 2012) and GATK (McKenna et al., 2010) for SNP calling; SPAdes (Bankevich et al., 2012), Velvet (Zerbino & Birney, 2008) and Flye (Kolmogorov et al., 2019) for *de novo* assembly; IQ-TREE 2 (Minh et al., 2020), RaxML-NG (Kozlov et al., 2019) and BEAST 2 (Bouckaert et al., 2019) for Phylogenetics; BLAST (Altschul et al., 1990), GSAlign (H.-N. Lin & Hsu, 2020), MUMmer4 (Marçais et al., 2018), and Minimap2 (Li, 2018) for comparative genomics.

**Expanding known genetic diversity of microbial pathogens**

The availability of genomic databases and the advancement of tools for their efficient analysis have highlighted that the characterized portion of the pathobiome is merely the tip of the iceberg. Indeed, new taxa, but also previously uncovered genetic diversity of known taxa are both frequently recovered using data mining approaches. A striking example is that, while only a few dozen viral species are currently reported to infect rice, an exhaustive study of all the 17,115 rice RNA sequencing runs available on NCBI SRA uncovered hundreds of new viruses (Zhu et al., 2025). Another example is the identification of over $10^5$ novel RNA viruses by searching for the hallmark gene RNA-dependent RNA polymerase in databases (Edgar et al., 2022). Identifying new taxa improves and expand the taxonomic classification of pathogens (Koonin & Lee, 2025; Reddy & Sidharthan, 2024; Rosani et al., 2023) but can also highlight pathogens presenting with a potential (Kawasaki Junna et al., 2021) or established (Yan et al., 2025) risk of emergence. Expanding genomic resources at different taxonomical levels also benefits our understanding of the evolutionary history of microbial pathogens. For example, a data mining-based study has suggested that gene exchange between Nidovirales (the order including *Coronaviridae*, a viral family that has recently received a lot of attention) might be more frequent than previously thought and may facilitate host jumps (Lauber et al., 2024). At the species level, ancient DNA analysis from neurotoxigenic *Clostridium tetani* in archaeological human samples revealed a subgroup from South America that produces an unknown tetanus neurotoxin variant (Hodgins et al., 2023). Research questions benefiting from expanded genomic resources of microbial pathogens are numerous, and listing them goes beyond the scope of this paper. They include the characterization of the structuring of the genetic diversity, variability in host/pathogen interactions, the lineage organization, the depiction of evolutionary events, genome organization, or delineation of the pangenome (for a review, see Vello et al., 2024). Beyond fundamental studies, the characterization of the intraspecific diversity is a crucial early step in the study of a pathogen, as it conditions the establishment of disease management and control strategies. For viral and bacterial pathogens, popular detection methods include various molecular and serological methods (Rajapaksha et al., 2019). Although their susceptibility to genetic and antigen diversity varies greatly, their development relies on a good knowledge of the pathogen's diversity to ensure sensitivity across all lineages. The same goes for vaccine development, which needs to target conserved antigens. Ultimately, while expanding genomic resources of microbial pathogens enhances both our fundamental understanding of pathogen evolution and our ability to anticipate and respond to pathogen emergence, the genomic resources brought by and to the scientific community through their sharing on public databases can be even more useful when accompanying metadata are present.

## *In silico*-based epidemiological surveillance : spatiotemporal distribution of microbial pathogens

Extensive efforts are dedicated to the characterization of the spatiotemporal distribution of pathogens. Initiatives such as EPPO (https://gd.eppo.int/) or CABI (https://www.cabidigitallibrary.org/) for plant pathogens, Atlas ECDC (http://atlas.ecdc.europa.eu/public/index.aspx), GISAID (https://gisaid.org) or Nextstrain (https://nextstrain.org/) for human pathogens aggregate occurrence data from multiple sources and aim to provide up-to-date distributions of pathogens. This knowledge is crucial to focus epidemiological surveillance efforts, but also to better understand the migration of pathogens, their emergence, and the factors driving it. Pathogen distributions are subject to ongoing changes due to new emergences –following the concept of biotic homogenization (Bebber et al., 2014)– and their detection and the reporting of newly infested territories based on direct observation can therefore experience some delays. Current distributions might be enriched by indirect occurrence data derived from genomic data mining, possibly uncovering previously undetected occurrences. This approach relies on the metadata accompanying genomic dataset. For example, this approach was used to expand the known geographic distribution of the cotton leafroll dwarf virus (Olmedo-Velarde et al., 2024) and the Solanum nigrum ilarvirus 1 (Rivarez et al., 2023). In the same manner, the date of isolation of the samples can be very informative, both directly by providing evidence of the presence of the pathogen at a given date (Olmedo-Velarde et al., 2024), or indirectly by enriching inferences of the past evolutionary history of pathogens (R. C. Ferreira et al., 2024). Authors have also exhaustively screened the SRA database to confirm the endemicity of a bacterial genus to New Zealand (Power et al., 2024). To finish, spatiotemporal data obtained through the genomic screening of public repositories represent a powerful complement to traditional epidemiological surveillance monitoring methods, but are also useful to enrich models of pathogen spread and evolution.

## Uncover ecological interactions: microbial communities, hosts and vectors

Microbial pathogens are involved in complex relationships with their biotic environment, which data mining approaches can help shed light on. By definition, microbial pathogens interact with the organisms they infect, but they sometimes also interact with vectors (most commonly insects) and microbial communities. Host range remains one of the key epidemiological characteristics of pathogens. Indeed, for example, more than half of human pathogens are zoonotic, and those having a broad host range are most likely to cause disease emergence (Woolhouse & Gowtage-Sequeria, 2005). As such, genomic data mining is often useful to define or expand pathogens' host range (including the identification of pathogen reservoirs), identify broad host range pathogens, uncover host jump events, or identify the genetic determinants of host specificity (Reddy & Sidharthan, 2024; Sidharthan et al., 2024; Thava Prakasa Pandian et al., 2024). Although the cases aforementioned identified the host component of the host-pathogen interaction using metadata, we speculate that the specificity of the interaction can be refined to the host genotype in cases where studying adaptive or resistance mechanisms is relevant. Querying sequencing databases with vectors or vector-borne pathogens can also enrich known pathogen-vector associations by identifying novel pathogens of known vectors (de Andrade et al., 2024; Y. Lin & Pascall, 2024) or by expanding the potential vector range of known pathogens (L. Y. M. Ferreira et al., 2024). Indeed, the vector range of vector-borne diseases is a key factor for disease management because control measures often target vectors rather than the transmitted pathogen (e.g., malaria, dengue, and most phytopathogenic viruses). Besides the pathogens they transmit, the characterization of the vectors' microbiome is also valuable to identify their own pathogens, which might potentially serve as biocontrol agents (de Andrade et al., 2024; Debat et al., 2024). For instance, Gupta et al. identified a novel virus likely infecting the neurotropic parasite *Toxoplasma gondii* through the screening of human neuronal genomic transcriptome datasets (Gupta et al., 2024). As ecological interactions involving microbial pathogens can be complex and numerous, increasing the sample size and the heterogeneity of a genomic dataset using data mining is a good strategy to deepen our understanding of pathogens' ecological interactions.

**Challenges in using public genomic data**

While genomic data recovered from data mining strategies are informative *per se* for genetic diversity studies for example, further studies often rely on the accompanying metadata. The most commonly provided –and used– information includes host, location, and date of isolation, but sequencing runs with all three fields filled are unfortunately scarce. When absent, information can be tediously gathered by manually going through the associated publications –a process that might soon benefit from automation, given the recent advances in natural language processing algorithms. When present, metadata must still be taken with caution, because several factors can make them misleading or even inaccurate. First, the date metadata field should represent the date of sampling rather than the date of sequencing, but special cases (e.g. "sampling" a leaf on a herbarium plant specimen, for example) can lead to date misattribution by the submitter. Depending on the study, flagging laboratory-maintained organisms might also be appropriate because the rate and directionality of their evolution might be specific to the artificial environment of the laboratory. Moreover, for pathogens, the relevant host species to consider can derive from the "host" metadata field or the "organism" metadata field, depending on whether the sequencing run is dedicated to the study of the pathogen or its host. Metadata inaccuracy can also stem from contamination, a term referring to genetic material within a sample that did not originate from that specific biological sample and to which the metadata therefore does not apply. External contamination comes from the sample's surrounding environment, including laboratory equipment, library preparation kits, sequencing platforms, and researchers themselves (Eisenhofer et al., 2019; Mahillon et al., 2024). Cross-sample contamination can originate from well-to-well contamination, index switching, or sample bleeding (Ballenghien et al., 2017; Lou et al., 2023; Vigne et al., 2018). Strategies employed to tackle this problem include specific laboratory practices, the addition of negative controls, and the *in silico* removal of contaminants (Eisenhofer et al., 2019). These approaches are unfortunately poorly adapted to the reuse of publicly available data (*a posteriori* implementation is mostly impossible), but also to the study of pathogens, which are often only represented at low frequencies among host or vector reads (complicating the *in silico* removal of contaminants). In some specific cases, it is possible to suspect cross-sample contamination and hence exclude the corresponding sequencing runs from a study. Indeed, as samples sequenced in the same batch often share the same BioProject accession on NCBI, the detection of the same genotype of a pathogen in two runs within a BioProject calls for a careful check. Given the difficulties in distinguishing contaminants from genuine presence in publicly available sequencing runs, one should remain cautious when interpreting results obtained from data mining studies. A good strategy is to consider these results preliminary and to confirm them with other data sources. This can be done by (i) checking that the results obtained rely on multiple independently generated datasets, (ii) making sure that the results obtained from data mining are compatible with their data mining-free corollary (for example, relying only on curated published genomes), (iii) collaborating with the authors of the relevant sequencing runs to resequence the sample if it still exists, or (iv) re-collecting similar samples with a dedicated study.

**Conclusion and perspectives**

Genomic data mining has proven to be a powerful approach to uncover previously unrecognized aspects of microbial pathogen biology, ecology, and evolution. By repurposing publicly available sequencing data, researchers can deepen our understanding of the epidemiology and the evolutionary history of microbial pathogens without generating new sequencing data. While the variability in quality, completeness, and accuracy of the associated metadata often prevents any conclusive study based solely on data mining, the approach represents a powerful springboard for hypothesis generation and preliminary analyses. Notably, databases remain biased towards well-studied taxa. Specifically, the scarcity of systematic and metadata-rich sequencing efforts at ecological scales continues to constrain the full potential of the approach. Filling this gap could open the door to addressing new challenges, such as going from the sole detection of the presence of a taxon to the understanding of its ecological role by uncovering the interactions it has with other organisms. Furthermore, genomic data can be combined with data from other fields of study. For example, geospatial data (e.g. altitude, land use, soil, atmospheric properties, degree of anthropization, etc.) can enhance our understanding of the effect of abiotic factors on pathogens; text mining data (screening of scientific articles, search engine queries, social network posts, wikipedia traffic, etc.) can provide spatiotemporal distributions of pathogens for epidemiosurveillance programs; physiological, phenotypic, or yield data can be useful to characterize the phenotype of a disease (incidence, severity, ...). To finish, the willingness of the researchers to make their data and associated metadata publicly available is a crucial enabler for data mining approaches and deserves commendation. Equally important is the collaborative work dedicated to database interoperability –as exemplified by initiatives like the International Nucleotide Sequence Database Collaboration– which has significantly amplified the usability of genomic databases. Looking ahead, broader adoption of the FAIR (Findable, Accessible, Interoperable, Reusable) data principles would enhance even further the utility of genomic data mining.

**Acknowledgments**

The authors thank Eugénie Hébrard (PHIM) for her thoughtful comments on the draft of the paper. This work has been publicly funded by the ANR (the French National Research Agency) under the project SPADYVA (ANR-20-CE35-0008-01).

**Data Availability**

No dataset was generated nor used for this review study.